\documentclass[aps,prl,twocolumn,preprintnumbers,amsmath,amssymb]{revtex4-1}
\usepackage{graphicx}
\usepackage{subfigure}
\usepackage{epsfig}
\usepackage{dcolumn}
\usepackage{bm}
\usepackage{ulem}
\usepackage{color}

\def\be{\begin{equation}}       \def\ee{\end{equation}}
\def\bea{\begin{eqnarray}}      \def\eea{\end{eqnarray}}
\def\ba{\begin{array}}
\def\ea{\end{array}}
\def\bnum{\begin{enumerate} }
\def\enum{\end{enumerate}}

\def\nn{\nonumber}

\def\=>{\Rightarrow}
\def\>{\rightarrow}

\def\PRB{Phys. Rev. B}
\def\PRL{Phys. Rev. Lett.}
\def\eye2{Fathbb{I}}

\def\Eq#1{Eq.~(\ref{#1})}

\def\Tr{\mathrm{Tr}}

\renewcommand{\>}{\rangle}

\begin{document}
\title{Correlated double-Weyl semimetals with Coulomb interactions: Possible applications to HgCr$_2$Se$_4$ and SrSi$_2$}
\author{Shao-Kai Jian$^1$}
\author{Hong Yao$^{1,2}$}
\email{yaohong@tsinghua.edu.cn}
\affiliation{$^1$Institute for Advanced Study, Tsinghua University, Beijing 100084, China \\
$^2$Collaborative Innovation Center of Quantum Matter, Beijing 100084, China}

\begin{abstract}
We study the fate of double-Weyl fermions in three-dimensional systems in the presence of long-range Coulomb interactions. By employing the momentum-shell renormalization group approach, we find that the fixed point of noninteracting double-Weyl fermions is unstable against Coulomb interactions and flows to a nontrivial stable fixed point with anisotropic screening effect. Moreover, experimentally measurable quantities such as specific heat obtain unusual logarithmic corrections. 
We also discuss implications of our results to three-dimensional materials HgCr$_2$Se$_4$ and SrSi$_2$, candidate materials of hosting double-Weyl fermions.
\end{abstract}

\date{\today}
\maketitle

\textbf{Introduction:} While characterizing the nature of possible exotic quantum phases in correlated electronic systems with large Fermi surfaces remains a central and challenging issue, electronic systems with only discrete Fermi points have attracted increasing attention in the past few years\cite{Vafek-14}. With only Fermi points, new phenomena beyond conventional Fermi liquids are often expected when appropriate short-range interactions or long-range Coulomb interactions are taken into account.
For instance, for systems with simple linear band crossing such as single-layer graphenes\cite{Castro-Neto-09} and surface states of topological insulators\cite{Hasan-10,XLQi-11} as well as three-dimensional Weyl semimetals, Lorentz invariance\cite{Vafek-14} and, under certain conditions, space-time supersymmetry\cite{SSLee-07,KunYang-08,Grover-14,SKJian-14} could emerge in low energy and long distance. For systems with nonlinear dispersions around band-touching points, the interaction effect is often more manifest due to enhanced density of states. The simplest such nonlinear dispersion is the case of quadratic band touching (QBT)\cite{KaiSun-09}. It was shown that for two-dimensional QBT systems spontaneous symmetry breaking already occurs at infinitesimal short-range interactions\cite{KaiSun-09,QingLiu-10,Vafek-10,Lemonik-10,WFTsai-11}.

For QBT points in three-dimensional (3D) cubic systems such as Pr$_2$Ir$_2$O$_7$, weak short-range interactions are simply irrelevant mainly because of vanishing density of states. However, long-range components of Coulomb interactions are relevant at the noninteracting fixed point and may induce a non-Fermi liquid\cite{Abrikosov-74,Moon-13} or a Mott insulator\cite{Herbut-14,HHLai-14}, which is in sharp contrast to marginally irrelevant Coulomb interactions\cite{Hosur-12,Isobe-12,Isobe-13,Goswami-11,WWK-14, HZ-Lu} in Weyl semimetals\cite{XGWan-11,Burkov-11,YBKim-12,HMWeng-15,BQLv-15,SYXu-15}. Another nontrivial band crossing whose dispersion is not simply linear is the double-Weyl point, around which spectrum disperses linearly only along one momentum direction but quadratically along the two remaining directions\cite{GangXu-11,ChenFang-12,SMHuang-15}. As suggested by its name, a double-Weyl point acts like a magnetic monopole in crystalline momentum space with twice the monopole charge of usual Weyl points. Double-Weyl fermions may be realized in materials breaking time-reversal or inversion symmetry and were predicted to occur in the candidate material HgCr$_2$Se$_4$\cite{GangXu-11} with two separated double-Weyl points which are protected by the material's crystalline $C_4$ symmetry. Weak short-range interactions are certainly irrelevant at the noninteracting fixed point of double-Weyl fermions. Nonetheless, similar to the QBT in 3D, the vanishing density of states for double-Weyl fermions may not screen the long-range Coulomb interactions effectively and one naturally asks if the long-range Coulomb interactions can induce exotic new physics in double-Weyl semimetals.

In this paper, we investigate the fate of double-Weyl fermions in the presence of long-range Coulomb interactions, particularly motivated by the candidate double-Weyl semimetal material HgCr$_2$Se$_4$\cite{GangXu-11} with two double-Weyl points. (Note that two is the minimum number of double-Weyl points because of the no-go theorem\cite{no-go}.) Here are the main findings. At the noninteracting fixed point of double-Weyl fermions, the isotropic part of the Coulomb interaction is marginally irrelevant while its anisotropy is relevant; the noninteracting fixed point is then not stable and the system flows into a stable nontrivial fixed point with anisotropic Coulomb screening. The anisotropic Coulomb screening is mainly due to the anisotropic dispersions of double-Weyl fermions. Compared with the linear-dispersing direction, the quadratic-dispersing directions have larger density of states which results in stronger screening. Specifically, the Coulomb interactions along the quadratic directions are screened to a faster-decaying $\frac1{r^2}$ behavior, while it is still $\frac1r$ along the linear-dispersing direction. At this fixed point with anisotropic Coulomb screening, the Coulomb interaction is again marginally irrelevant, which is in sharp contrast to the anisotropic Coulomb screening in anisotropic Weyl fermions at whose fixed point the Coulomb interaction is eventually irrelevant\cite{BJYang-14}. Because of the marginally irrelevant Coulomb interactions at the stable fixed point, various physics quantities receive logarithmical corrections and we explicitly analyzed the temperature dependence of specific heat and found exotic logarithmical corrections at low temperature.

\textbf{Model:} We start with a simple microscopic model of spinless fermions (namely, spin-polarized electrons close to the Fermi level of the ferromagnetic half metal HgCr$_2$Se$_4$), which hosts two double-Weyl fermions. We consider the following effective model featuring two double-Weyl fermions:
\bea\label{lattice-model}
H_0&=&\sum_{\vec k} c^\dag_{\vec k} \Big\{ 2t_1 (\cos k_1-\cos k_2)\sigma_1 + t_2 \sin k_1 \sin k_2 \sigma_2 \nn\\
&&~+ 2t_3 (\cos k_1+\cos k_2+\cos k_3 - m)\sigma_3\Big\} c_{\vec k},
\eea
where $c^\dag_{\vec k\sigma}$ creates a spin-polarized electron with orbital index $\sigma=1,2$ and $t_j$ are hopping parameters. We assume that the system respects $C_4$ rotational symmetry along the $z$ axis as well as the mirror symmetry $R_z$ changing $z\to -z$. The phase diagram as a function of $m$ is shown in Fig. 1. We shall focus on the region $1<m<3$ where two double-Weyl points are realized at $\vec Q_\pm= (0,0,\pm k^\ast_3)$ with $k^\ast_3=\arccos(2-m)$. Specifically, the $C_4$ symmetry protects the double-Weyl points, while the $R_z$ symmetry requires that the two double-Weyl points have equal energy.

Because of vanishing density of states at double-Weyl points, long-range Coulomb interactions cannot be screened to short-range ones. Here, for simplicity, we consider instantaneous $1/r$ Coulomb interactions between electrons, which is a good approximation to the retarded one since the speed of light in materials is much faster than the Fermi velocity of electrons. By Hubbard-Stratonovich transformation, we introduce
a bosonic field to decouple four-fermion Coulomb interactions so that we consider the following effective action to study the fate of double-Weyl fermions in the presence of Coulomb interactions:
\begin{eqnarray}
S &=& \int d^4 x \bigg\{ \bar{\psi} \Big{[} (\partial_0 +ig\phi)\Gamma_0 - \frac{i}{2m} d_j(-i\vec \nabla)\Gamma_j  \nonumber\\
&-& iv(-i\partial^3)\Gamma_3 \Big{]} \psi + \frac{1}{2} \Big[( \partial_1\phi)^2 +( \partial_2\phi)^2 +\eta (\partial_3 \phi)^2\Big]  \bigg\}, ~~~ \label{action}
\end{eqnarray}
where $\Gamma_0=\tau_1,\Gamma_1=\sigma_1\tau_2,\Gamma_2=\sigma_2\tau_2, \Gamma_3=\sigma_3\tau_2$, $d_1(\vec k)=k_1^2-k_2^2, d_2(\vec k)=2k_1 k_2$, $\tau_j$, and $\sigma_i$ are Pauli matrices representing valley space and orbital space, and $\psi^\dag=(\psi^\dag_+,\psi^\dag_-)$ where $\psi^\dag_\pm$ represents the low-energy electron near two band-touching points $\vec Q_\pm$. Note that we have implicitly assumed above that the coupling between different double-Weyl fermions is irrelevant when the separation of double-Weyl points is large compared to the momentum scale that we are interested in (see Appendix). The Coulomb interaction is proportional to $g^2/v$. In the above action, we also introduced $\eta$ to represent the anisotropy of Coulomb interactions which is allowed by the crystalline symmetry of the system. As shown below, $\eta$ flows to zero in the infrared fixed point which renders qualitatively different Coulomb interactions in the $xy$ plane and along the $z$ axis.

\begin{figure}
\includegraphics[width=8.4cm]{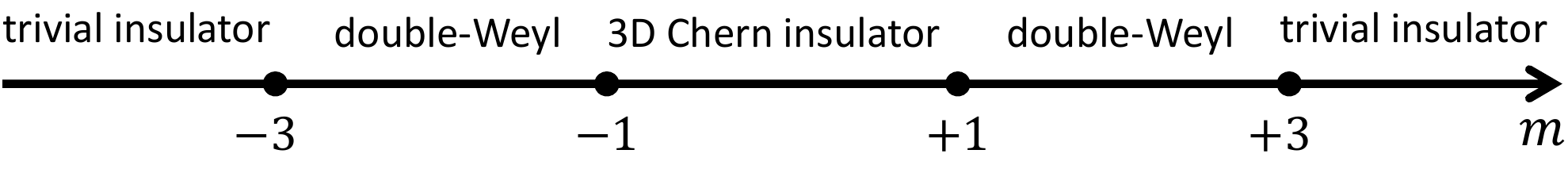}
\caption{The schematic phase diagram of the microscopic lattice model in \Eq{lattice-model} as a function of the parameter $m$. Here, ``3D Chern insulator'' refers to a 3D insulator with finite Hall conductivity $\sigma_{xy}$. Quadratic band touching is realized at the transitions between the double-Weyl semimetal and the trial (or Chern) insulators.}
\label{phase}
\end{figure}

\textbf{Screening from RPA analysis:} It is well known that long-range Coulomb interactions between electrons are screened by low-energy electrons and become short-ranged in materials with large Fermi surface. Nevertheless, in a double-Weyl semimetal, the Fermi surface shrinks into points and the density of states vanishes at zero temperature so that the screening effect needs to be reconsidered. In this section, we implement a random phase approximation (RPA) calculation to investigate the screening effect in a double-Weyl semimetal. The low-energy effective action is given by Eq. (\ref{action}). In standard RPA calculations, fermions are integrated out to get electron-hole polarization, the quantum correction to Coulomb interactions. A straightforward calculation shows (see Appendix for details) that electron-hole polarization is anisotropic,
	\bea
	 	-\Pi(p) & \propto& p_\perp^2+|p_z|, \label{RPA_potential}
	\eea
where for simplicity we have implicitly neglected insignificant coefficients as well as a potentially-unimportant logarithmic factor before $p_\perp^2$. Since the linear part of electron-hole polarization is more important than the quadratic part in the infrared, the higher order of $p_z$ can be omitted at the long-wavelength limit so that we obtain the screened Coulomb potential $V_\text{C}(p)\sim \frac{1}{p_x^2+p_y^2+|p_z|}$. This means that the bare isotropic Coulomb interaction becomes highly anisotropic due to the screening by the double-Weyl fermions. In real space, the Coulomb potential along the $z$ axis is qualitatively unchanged, {i.e.}, $V_\text{C}(0,z) \propto \frac{1}{|z|}$, while in $xy$ plane, the Coulomb potential behaves as $V_\text{C}(r_\perp,0) \propto \frac{1}{r_\perp^2}$; namely, it decays qualitatively faster due to the screening effects of the low-lying quadratic fermions in the $xy$ plane.

\textbf{Renormalization group analysis:} In this section, we implement a momentum-shell renormalization group (RG) analysis to show that rather than being screened all together, Coulomb interaction is marginally irrelevant in double-Weyl semimetals at low energy which implies that various measurable physical quantities receive logarithmic corrections at low temperature or low energy.

Since the double-Weyl fermions disperse with different power behaviors along different momentum directions, the scaling dimensions of spatial coordinates can be different. So, without loss of generality, we assume that scaling dimensions are given by $[x_3]=-z_3,[x_1,x_2]=-z_1,[t]=-1$. In the following RG analysis, we integrate out the degree of freedoms in a momentum shell of $\Lambda e^{-l}<E_{\vec k}<\Lambda$, where $l>0$ is the flow parameter, $\Lambda$ is the ultraviolet energy cutoff, and $E_{\vec k}=\sqrt{\frac{(k_x+k_y)^4}{4m^2}+v^2 k_z^2}$ which is the energy of double-Weyl fermions. We emphasize that momentum shell of this kind correctly includes the high-energy degrees of freedom which effect the low-energy physics. From the various coupling constants in the action, we construct two dimensionless coupling constants that will be useful to characterize the RG flow: $\alpha=\frac{g^2}{8\pi^2 v}$ and $\beta=\frac{g^2mv}{8\pi \Lambda\eta}$. Note that the dimensionless parameter $\alpha$, which has the form of a fine-structure constant, characterizes the strength of Coulomb interactions.

\begin{figure}
\subfigure[]{\vspace{-0.8cm}\includegraphics[width=2.4cm]{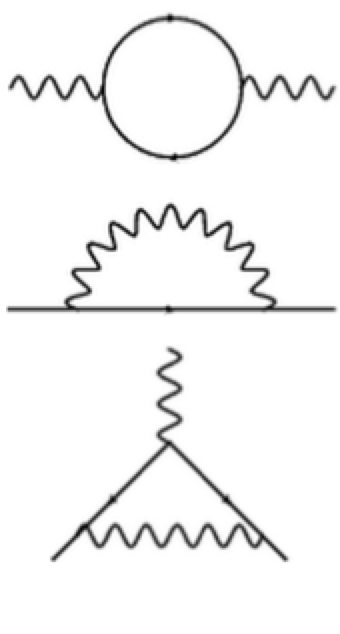}}~~~
\subfigure[]{\includegraphics[width=5.2cm]{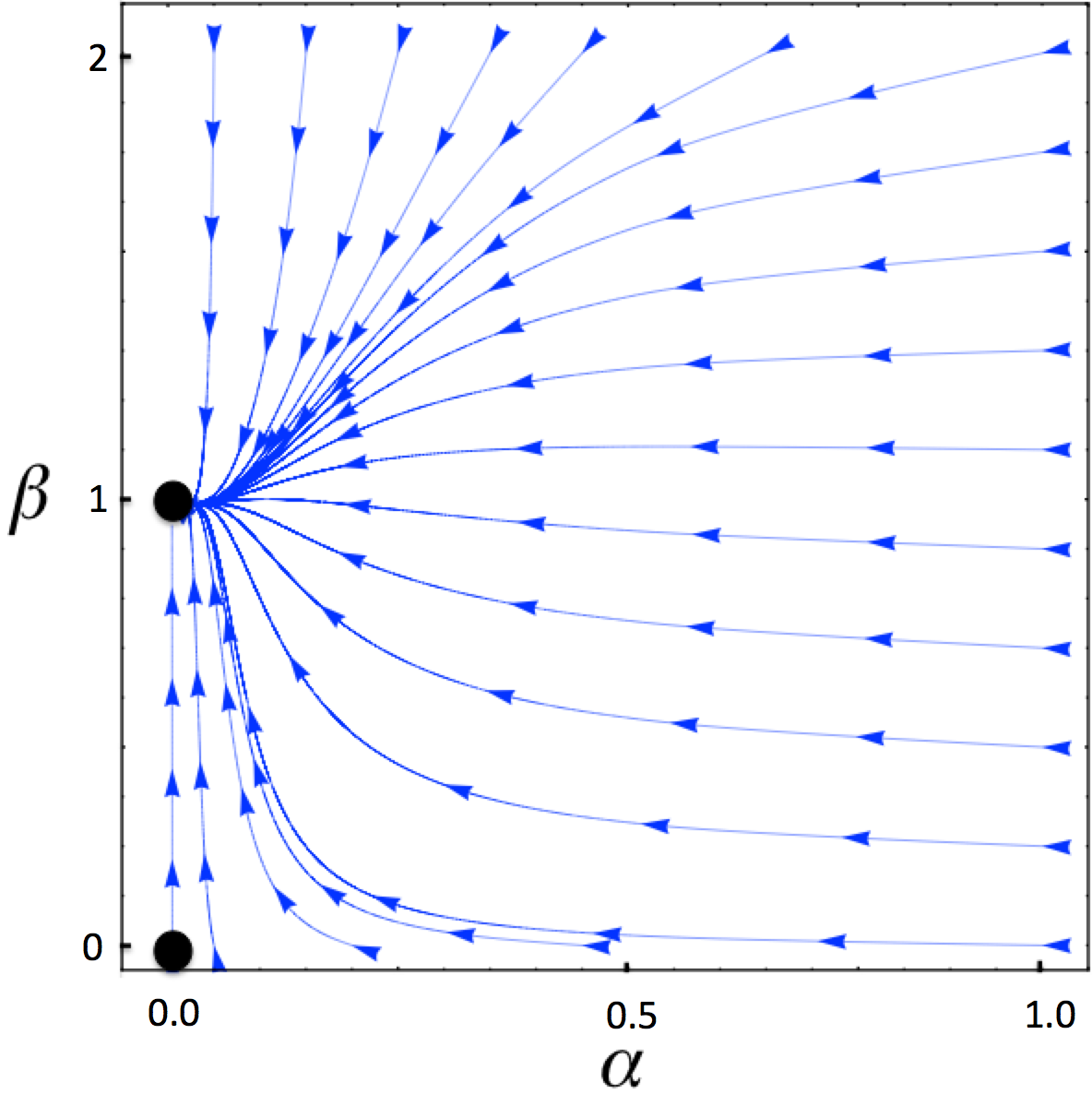}}
\caption{(a) Feynman diagrams. (b) The RG flow in the parameter space of $\alpha=\frac{g^2}{8\pi^2 v}$ and $\beta=\frac{g^2mv}{8\pi \Lambda\eta}$. There are two fixed points. The fixed point at (0,0) represents noninteracting double-Weyl fermions which are not stable against long-range Coulomb interactions. The nontrivial fixed point is at $(\alpha^\ast,\beta^\ast)=(0,1)$, where the Coulomb interactions at low energy are qualitatively different between the quadratic- and linear-dispersing directions. }
\label{rg_flow}
\end{figure}

We outline the main results of RG analysis here and provide calculational details in Appendix. We choose $z_1$ and $z_3$ such that the physical quantities $m$ and $v$ characterizing the dispersion of double-Weyl fermions are finite at the fixed point. Then, we can use only two parameters $\alpha$ and $\beta$ to depict the full RG flow. By  iterating the RG equations, we obtain the RG flow of $\alpha$ and $\beta$, as shown in Fig. \ref{rg_flow}(b).
It is clear that there are only two fixed points: $(\alpha^\ast,\beta^\ast)=(0,0)$ is the unstable fixed point of noninteracting double-Weyl fermions while $(\alpha^\ast,\beta^\ast)=(0,1)$ is the stable one. Near the fixed points, the RG equations to the lowest order of $\alpha$ and $\beta$ are given by
 	\bea
	\frac{d \alpha}{ dl} &=& -\frac{19}{3} \alpha^2,  \\
	\frac{d\beta}{dl} &=& (1-\beta+\alpha)\beta,
	\eea		
from which it is apparent that the Coulomb interaction characterized by $\alpha$ is marginally irrelevant close to the stable fixed point of $(0,1)$. It is known that marginally irrelevant interactions will give rise to logarithmic corrections to physical quantities in low energy or low temperature. Besides being marginally irrelevant, Coulomb interaction becomes infinitely anisotropic due to the screening of anisotropic gapless double-Weyl fermions, namely, $\eta\propto \alpha/\beta \to 0$ in the infrared. This phenomenon is captured by the particle-hole polarization,
    \bea
    p_\perp^2+\eta p_z^2-\Pi(p) \sim p_\perp^2(1+\frac{16}{3\pi}\alpha l) +\eta p_z^2(1+\beta l).
    \eea
At the stable fixed point $(\alpha^\ast,\beta^\ast)=(0,1)$, we find that boson kinetic energy in the $xy$ plane gets no qualitative renormalization, while it is strongly renormalized along the $z$ axis. This renormalization can be understood as anomalous dimensions,
	\bea
	\vec{p}^2-\Pi(p) &=& p_x^2+p_y^2+p_z^2[1+\ln \frac{\Lambda}{v |p_z|} ]  \nonumber\\
	&\sim& p_x^2+p_y^2+|p_z|.
	\eea
In other words, the Coulomb potential is renormalized to be $V_\textrm{C}(\vec p)=\frac{1}{p_x^2+p_y^2+|p_z|}$. This result is qualitatively consistent with the one from the RPA calculation in Eq. (\ref{RPA_potential}), which we believe further shows that the divergent anisotropic screening of the Coulomb interaction is correct.

\textbf{Screening of a Coulomb impurity:} We expect that the anisotropic dispersion of double-Weyl fermions can cause the unusual feature of induced charge density of a Coulomb impurity, which interacts with the double-Weyl fermions through usual Coulomb potential proportional to $1/r$. We shall consider both the case of noninteracting double-Weyl fermions and of Coulomb-interacting double-Weyl fermions. For noninteracting double-Weyl fermions, a Coulomb impurity causes a divergent distribution of density of induced charges, while for Coulomb-interacting double-Weyl fermions, the induced charge distribution is instead a simple $\delta$ function located around the impurity. This $\delta$ function feature of induced charge density is caused by effective screening of the Coulomb interactions by double-Weyl fermions.

To be specific, we consider an impurity of charge $Ze$ located at the origin, where $Z$ is an integer constant. For noninteracting double-Weyl fermions, this impurity generates a bare Coulomb potential to them, namely, $V_0(\vec q)= \frac{g_0^2}{|\vec q|^2}$, where $g_0$ characterizes the Coulomb interaction between the impurity and the double-Weyl fermions. Then, the density of induced charges is given by $\rho_0(\vec r)=\int \frac{d^3 q}{(2\pi)^3} \rho_0(\vec q)$, where $\rho_0(\vec q)= Ze V_0(\vec q) \Pi(\omega=0,\vec q)$, with $\Pi(\omega,\vec q)$ being the particle-hole polarization which we assume,  without loss of generality, to be $\Pi(q)=-B_\perp q_\perp^2-B_z |q_z|$ according to RPA calculations. The anomalous characteristic of charge distribution can be seen from partially induced charge densities, i.e., $Q_z(z)= \int d^2 r_\perp \rho(\vec r)$ and $Q_\perp(r_\perp)=\int dz \rho(\vec r)$ with $Q_\perp(r_\perp) \propto  \delta^{(2)}(r_\perp)$ and $Q_z(z) \propto \log |z|+\textrm{const}$. Note that partially induced charge density $Q_z$ above is divergent at the long-wavelength limit, which arises because of the particular dispersion of noninteracting double-Weyl fermions. Such singularity does not appear when we consider Coulomb interactions in the double-Weyl fermions.

For the interacting case, the Coulomb interaction is renormalized as $V(\vec q)^{-1}= q_x^2+q_y^2+|q_z|$, where we have rescaled momentum to make the parameters look simpler. The induced charge density is now given by
    \bea
	\rho(\vec q)&=& Ze V(\vec q) \Pi(\omega=0,\vec q) \nn\\
    &=& -\frac{Ze(B_\perp q_\perp^2+B_z |q_z|)}{q_x^2+q_y^2+|q_z|},
    \eea
from which we can compute $Q_z(z)$ and $Q_\perp(r_\perp)$.
Furthermore, $\int dz Q_z(z) \equiv \int d^2 r_\perp Q_\perp(r_\perp)$ requires that $B_\perp=B_z$ (see Appendix). Consequently, we obtain $\rho(\vec q) = B_z Ze$, where $B_z$ is a constant of order one. It is now clear that $\rho(\vec r)=B_zZe\delta(\vec r)$, where $\delta(\vec r)$ is the usual $\delta$ function. This feature is similar to the induced charge discussed in Ref. \cite{BJYang-14}, though the Coulomb interaction there is irrelevant.

\textbf{Logarithmic corrections to specific heat:} Since the Coulomb interaction is marginally irrelevant at the nontrivial stable fixed point of $(0,1)$, measurable physical quantities will receive logarithmic corrections, which originate from the logarithmic behavior of the coupling constant under RG flow. More specifically, consider the fine-structure constant as a function of energy scale $E$,
	\begin{eqnarray}
	\alpha(E)= \frac{\alpha_0}{1+\frac{19\alpha_0}{3}\log \frac{\Lambda}{E}}, \label{fine_structure}
	\end{eqnarray}
where $\alpha_0$ is the bare fine-structure constant and $\Lambda$ is the ultraviolet energy cutoff. It is clear that as the energy/temperature decrease, the interaction strength becomes weaker in a logarithmic manner. One would expect that this feature shall bring logarithmic corrections to various physical quantities.

Near the fixed point, the free-energy density scales as $f(T)= b^{-2z_1-z_3-1} f_0(\Lambda)$, where $b \equiv \frac{\Lambda}{T}$ and $z_1,z_3$ are scaling dimensions of the $xy$ coordinates and $z$ coordinate, respectively\cite{Schmalian-07}. For the specific heat $C(T)=-T \frac{\partial^2 f}{\partial T^2}$, we obtain $C(T)= b^{-2z_1-z_3} C_0(\Lambda)$.  Remember that we have  set the scaling dimension of time to be one, while allowing $z_1,z_3$ to flow. As discussed earlier, close to the fixed point, we choose $z_1=\frac{1}{2}(1-c_1\alpha\sqrt{\beta}-\sqrt{\frac{\beta}{4\pi}}\alpha\ln\alpha)$ and $z_3=1+\frac{\alpha}{\pi}-\frac{\pi \alpha^2}{4}$, where $c_1\approx0.355$, such that the mass $m$ and Fermi velocity $v$ in the dispersion of double-Weyl fermions remain fixed. Taking account of the logarithmic flow of fine-structure constant $\alpha$, the specific heat at low temperature (for details see Appendix) is given by
    \bea
	C(T)
	&\propto& \frac{T^2}{(1+\frac{19\alpha_0}{3}\ln \frac{\Lambda}{T})^{\frac{3c_2}{19}}} \nonumber\\
	&&\times \exp\left[ -\frac{3}{76\sqrt{\pi}}\left( \ln(1+\frac{19\alpha_0}{3}\ln \frac{\Lambda}{T} )\right)^2
    \right], ~~\label{second}
    \eea
where $c_2=c_1-\frac1\pi \approx 0.037$. The first line is the usual logarithmic correction to specific heat found in much of the literature. However, the second line in Eq. (\ref{second}), which actually dominates at low energy, is a different phenomenon that we find for double-Weyl fermions with marginally irrelevant Coulomb interactions. It decays faster than the usual logarithmic correction, which indicates that this marginal phenomenon deviates from Fermi liquid in a more radical way and may be tested in future measurements of HgCr$_2$Se$_4$ and SrSi$_2$ at low enough temperature. It is crucial for the Fermi level to be at the double-Weyl points to observe the above logarithmic behavior of specific heat (see Appendix for a schematic specific heat curve). An interesting specific heat measurements in HgCr$_2$Se$_4$ was reported recently\cite{Wang}. However, because the Fermi level in the measured material deviates from double-Weyl points, electron contribution to the specific heat does not show the logarithmic correction in Eq. (\ref{second}). To possibly see the logarithmic corrections, it is desired to experimentally tune the Fermi level to the double-Weyl points in these candidate materials in the future. As discussed in Appendix, this unusual logarithmic correction is caused by the logarithmic term in the RG equations close to the fixed points. Logarithmic terms also appear in RG equations describing certain quantum criticality such as nematic quantum critical points in nodal $d$-wave superconductors\cite{Sachdev-08,Kivelson-08}.

\textbf{Discussions:} So far, we have considered a minimal model with only two double-Weyl points, {e.g.} in HgCr$_2$Se$_4$, interacting through Coulomb interactions. For such systems including HgCr$_2$Se$_4$, the Coulomb interactions are screened anisotropically: the screened Coulomb potential is qualitatively unchanged along the $z$ axis, { i.e.}, $V_\text{C}(0,z) \propto \frac{1}{|z|}$ but behaves as $V_\text{C}(r_\perp,0) \propto \frac{1}{r_\perp^2}$ in the $xy$ plane, which decays qualitatively faster because of screening by the low-lying quadratic fermions in the $xy$ plane. In a material that possesses many double-Weyl nodes along various directions, {e.g.}, in the predicted candidate material SrSi$_2$, one needs to take into account the contributions to particle-hole polarization from all double-Weyl fermions while the interactions between different double-Weyl fermions can still be neglected as long as we care about low-energy and long-wavelength physics, namely, $\Pi(p)=\sum_i \Pi_i(p)$, where $i$ ranges over all double-Weyl points. In general, there is no special spatial direction when many double-Weyl points are present. Thus, by averaging, we expect $\Pi(p) \propto |\vec p|$. In other words, the screened Coulomb potential has the form $V_\textrm{C}(\vec p) \sim \frac{1}{|\vec p|}$, where $|\vec p|=\sqrt{p_x^2+p_y^2+p_z^2}$. In real space, it behaves like $V_\textrm{C} ( r) \sim \frac{\Lambda J_1(r/a)}{r}$, where $J_1(x)$ is a Bessel function of the first kind and $a$ is of the order of the lattice constant. For $x\gg 1$, $J_1(x)\sim \frac{1}{\sqrt{x}}$. Thus, the Coulomb potential is modified to be $V_\textrm{C}( r) \sim r^{-3/2}$.
	
To conclude, in this work, we explored the fate of double-Weyl semimetals in the presence of long-range Coulomb interaction. Instead of being screened to be short ranged, the Coulomb interaction is marginally irrelevant at low energy. At the infrared stable fixed point, the Coulomb potential is strongly anisotropic, which is qualitatively different from the noninteracting Gaussian fixed point. Because of the marginal irrelevance of Coulomb interactions at the stable fixed point, physical quantities received logarithmic corrections in low energy or in low temperature. For instance, we find that the specific heat features an exotic logarithmic correction, deviating from the usual Fermi liquid, which originates from the interplay of anisotropic double-Weyl fermions and the bare isotropic interactions between them.

{\it Acknowledgement}: We would like to thank Xin Dai for helpful discussions. This work is supported in part by the National Thousand-Young-Talents Program (H.Y.) and by the NSFC under Grant No. 11474175 (SKJ and HY). {\it Note added}: In preparing the present work, we noticed an interesting work on a similar topic\cite{HHLai-15}. The RG scheme employed in our paper is different from Ref.\cite{HHLai-15} and some results are qualitatively different.

\begin{widetext}
\section{Appendix}
\renewcommand{\theequation}{A\arabic{equation}}
\setcounter{equation}{0}
\renewcommand{\thefigure}{\arabic{figure}}
\setcounter{figure}{0}

\subsection{1. Hubbard-Stratonovich transformation for Coulomb interaction}

	Using Hubbard-Stratonovich transformation, we introduce a boson field to decouple Coulomb interaction. In double-Weyl semimetal, low-energy excitations are described by gapless fermions around the two double-Weyl points. Thus, with a momentum cutoff near band-touching points, we can write fermion operators as
\bea
	 \Psi(\vec{r}) &=& \int \frac{d^3 k}{(2\pi)^3} \Psi(\vec{k}) e^{i\vec{k}\cdot\vec{r}} \nn\\
                   &=& \int_\Lambda \frac{d^3k}{(2\pi)^3}\Psi_Q(\vec{k}) e^{i\vec{k}\cdot\vec{r}} e^{i\vec{Q}\cdot\vec{r}}+ \int_\Lambda \frac{d^3 k}{(2\pi)^3}\Psi_{-Q}(k) e^{i\vec{k}\cdot\vec{r}} e^{-i\vec{Q}\cdot\vec{r}} \nonumber \\
	 &\equiv& \psi_Q(\vec{r}) e^{i\vec{Q}\cdot\vec{r}}+\psi_{-Q}(x) e^{-i\vec{Q}\cdot\vec{r}},
\eea
where the electron's orbital indices are implicit and $\Psi_{\pm Q}(k)=\Psi(k\pm Q)$. Then, the charge density operator can be expressed by these low energy fermions:	
\bea
	\rho(\vec{r}) &=& \Psi^\dag(\vec{r}) \Psi(\vec{r}) \nn\\
    &=&\rho_Q(\vec{r})+\rho_{-Q}(\vec{r}) + \left[\psi^\dag_{-Q}(\vec{r}) \psi_Q(\vec{r}) e^{i2\vec{Q}\cdot\vec{r}} +h.c.\right],
\eea
where $\rho_{\pm Q}(\vec r)$ represent contribution from double-Weyl fermions around band-touching points. There are also mixing terms with nonzero momentum. Then, the (Euclidean) action for the instantaneous Coulomb interaction between electrons is given by
\bea
	S_c&=& \frac{g^2}{2} \int d^3r d^3r' \frac{\rho(\vec{r})\rho(\vec{r}')}{4\pi |\vec{r}-\vec{r}'|} \nn\\
    &=& \frac{g^2}{2} \int d^3r d^3r' \frac{[\rho_Q(\vec{r})+\rho_{-Q}(\vec{r})][\rho_Q(\vec{r}')+\rho_{-Q}(\vec{r}')]}{4\pi |\vec{r}-\vec{r}'|} + \text{ finite momentum terms},
\eea	
where $g^2=e^2/\epsilon$ describes the strength of Coulomb interaction ($\epsilon$ is the dielectric constant of the material in question). Note that finite momentum terms behave like  $\int d^3r \frac{e^{i2 \vec Q\cdot \vec r}}{|\vec r|}$; for a generic finite momentum $\vec Q$, these terms are not important thanks to the fast oscillation factors $e^{ i2\vec Q \cdot\vec r}$ which are neglected in our calculations. Using a Hubbard-Stratonovich transformation, we introduce a boson field $\phi(x)$ to decouple the Coulomb interaction, namely,
\bea
    S_c &=& \int D\phi \exp \int d^4 x\left[ -\frac{1}{2} (\nabla \phi)^2 - i g \phi ({\psi}^\dag_{Q}
    \psi_Q+{\psi}^\dag_{-Q} \psi_{-Q})  \right].
\eea
We emphasize that boson field is not dynamic since it describes an instantaneous Coulomb interaction. This Coulomb potential plus a band Hamiltonian for double-Weyl fermion gives the effective action mentioned in main text.

\subsection{2. Random phase approximation}

Coulomb interactions can be screened to some intent by the low-energy fermions near double-Weyl points. The screening effect by low-energy electrons is given by the boson's self-energy appeared in RG analysis. Implementing random phase approximation (RPA), we shall consider only one loop diagram, which is first order quantum correction to boson self-energy (see next section):
\bea
	\Pi(p) &=& g^2 \int \frac{d^4 k}{(2\pi)^4} \Tr[\Gamma_0 G(k) \Gamma_0 G(k+p)] \nonumber\\
	&=& 4g^2 \int_k \frac{-w(w+\Omega)+(1/2m)^2(d_1(k)d_1(k+p)+d_2(k)d_2(k+p))+v^2
    k_3(k_3+p_3)} {(w^2+(1/2m)^2(d_1^2(k)+d_2^2(k))+v^2k_3^2)((w+\Omega)^2+(1/2m)^2(d_1^2(k+p)+d_2^2(k+p))+v^2(k_3+p_3)^2)},~~~~~
\eea
where $p=(\omega,\vec p)$, $\int_k\equiv \int \frac{d^4 k}{(2\pi)^4}$, and $d_1(k)=k_1^2-k_2^2$, $d_2(k)=2k_1k_2$. At this rate, we employ Feynman parameters and dimensional regularization to evaluate this integral (rescaling the momentum to absorb coupling constants)
\bea
	\Pi(p) &=& 4 g^2 \frac{2m}{v} \int_k \frac{-w(w+\Omega)+ k_3(k_3+p_3)+d_1(k)d_1(k+p)+d_2(k)d_2(k+p)}{(w^2+
    k_3^2+(k_1^2+k_2^2)^2)[(w+\Omega)^2+(k_3+p_3)^2+ ((k_1+p_1)^2+(k_2+p_2)^2)^2]}  \nn\\
    &=& \frac{2 g^2}{2\pi}\frac{2m}{v} \int_{x,\vec k_\perp} \frac{-x(1-x)(-\Omega^2+p_3^2)+d_1(k)d_1(k+p)+d_2(k)d_2(k+p)}{x(1-x)(\Omega^2+p_3^2) + (1-x)k_\perp^4+x|\vec k_\perp+\vec p_\perp|^4 },
\eea
where $x$ is the Feynman parameter, $\int_x\equiv \int_0^1 dx$, and $\vec k_\perp \equiv(k_1,k_2)$. It is clear that $\Pi(\Omega=0,\vec p=0)=0$ which is correct because of the vanishing density of states at the double-Weyl points. In the limit of low-energy, the momentum transfer is small so that we can obtain the polarization approximately. Momentum dependence of self-energy along $p_3$ direction are screened by the quadratic dispersion of fermion spectrum:
	\begin{eqnarray}
	\Pi(\Omega=0,\vec p_\perp=0,p_3) &=& \frac{2 g^2}{2\pi}\frac{2m}{v} \int_{x,k_\perp} \frac{-x(1-x)p_3^2+ k_\perp^4}{x(1-x)p_3^2
    + k_\perp^4},\nn \\
    &=&\frac{2 g^2}{2\pi} \frac{2m}{v} \int_{k_\perp} \frac{8k_\perp^4 \tanh^{-1} \frac{p_3}{\sqrt{4k_\perp^4 +p_3^2}}}{p_3\sqrt{4k^4_\perp+p_3^2}}  \nonumber \\
	&\propto& -|p_3|.
	\end{eqnarray}
The dependence on $p_x$ and $p_y$ is given by:
    \begin{eqnarray}
	\Pi(\Omega=0,\vec p_\perp,p_3=0)
    &=&  \frac{2 g^2}{2\pi} \frac{2m}{v} \int_{x,\vec k_\perp}
    \frac{(k_1^2-k_2^2)((k_1+p_1)^2-(k_2+p_2)^2)+4k_1
    k_2 (k_1+p_1)(k_2+p_2)}{(1-x)k_\perp^4+x(k_\perp+p_\perp)^4 },  \nonumber \\
    &\propto& -p_\perp^2,
    \end{eqnarray}
where we have implicitly neglected a potentially-unimportant logarithmic factor $\log p_\perp$. Consequently, we obtain electron-hole polarization:
	\begin{eqnarray}
	 \Pi(\vec p)&\propto& -(p_\perp^2 +|p_3|).
	\end{eqnarray}
We note that the momentum dependence of Coulomb interaction along the $p_3$ direction is strongly screened by electron-hole polarizations. Since at low energy, linear behavior of momentum $p_3$ is more important than bare quadratic momentum dependence, this will dramatically affect the Coulomb interaction: it becomes infinitely anisotropic. This anisotropy can be seen more directly in real space. The Coulomb interaction along $z$ axis is like usual $V(0,x_3) \propto \frac{1}{|z|}$; however, in the $xy$ plane, it behaves as $V(0,x,y) \propto \frac{1}{x^2+y^2}$. The effective interaction in the $xy$ plane after screening is qualitatively weaker because quadratic dispersions in the $xy$-directions render larger density of states at low energy and can screen the interactions more effectively.

\subsection{3. Renormalization group analysis}
Here we perform renormalization group (RG) analysis in details. The effective (Euclidean) action describing two double-Weyl fermions with long-range Coulomb interactions is
	\begin{eqnarray}
	 S &=& \int d^4 x \left\{ \bar{\psi}\left[ (\partial_0 +ig\phi) \Gamma_0 - \frac{i}{2m} d_j(-i\nabla)\Gamma_j - iv(-i\partial_3)\Gamma_3 \right] \psi+ \frac{1}{2} (( \partial_1\phi)^2 +( \partial_2\phi)^2 ) +\eta (\partial_3 \phi)^2 )\right\} ,
	 \end{eqnarray}
where the ultraviolet cutoff is implicit in the above equation. We can construct two dimensionless parameters from various coupling constants in the action above $\alpha=\frac{g^2}{8\pi^2 v}$ and $\beta=\frac{g^2mv}{8\pi \Lambda\eta}$. 	
In Wilsonian RG, integrating out the high-energy modes with $\mu<E_{\vec k}<\Lambda$, where $E_{\vec k}=\sqrt{[\frac{1}{2m}(k_1^2+k_2^2)]^2+v^2 k_3^2}$, will generate an effective action with lower energy cutoff. We derive the RG flow from iteratively integrating momentum shell whose fermions having energy $E_{\vec k}\in(\Lambda e^{-l},\Lambda)$, where $l\ge 0$ is the RG running parameter.  At the one-loop level, the effective action reads
	\begin{eqnarray}
	 S_{\text{eff}}= S_< - \frac{1}{2} \langle S_c^2 \rangle +\frac{1}{6} \langle S_c^3 \rangle,
	 \end{eqnarray}
where $S_c$ is the boson-fermion coupling which involves high energy modes in the momentum shell and $\langle \cdots \rangle$ means taking expectation value from high energy part of field configurations. There are three relevant contributions, {i.e.}, boson self-energy, fermion self-energy and vertex correction. It is worth noting that since the boson field is not dynamical, fermions do not get an anomalous dimension. Moreover, vertex correction at one-loop level vanishes. This is related to Ward identity.
	
It is straightforward to obtain particle-hole polarization and fermion self-energy through direct calculations as follows:
	\begin{eqnarray}
	\Pi(k) &=& g^2 \int \frac{d^4 q}{(2\pi)^2} \Tr[ \Gamma_0 G(q) \Gamma_0 G(q+k) ] \nn\\
    &=& -\frac{2g^2 l}{6\pi^2v} \frac{k_1^2+k_2^2}{2}- \frac{g^2mvl}{8\pi \Lambda\eta} \frac{\eta}{2} k_3^2,\\
	\Sigma(k) &=& -g^2 \int \frac{d^4 q}{(2\pi)^2} \Gamma_0 G(q) \Gamma_0 D(k-q) \nn\\
    &=& - \frac{g^2 x F_1(x)l}{8\pi v} v k_3\Gamma_3 - \frac{g^2 x F_2(x)l}{8\pi v} \frac{1}{2m} \left[ (k_1^2-k_2^2) \Gamma_1 + 2 k_1 k_2 \Gamma_2 \right],
	\end{eqnarray}
where $x\equiv \frac{\Lambda\eta}{mv^2}$ and $F_1,F_2$ are two positive functions whose behaviors will be discussed below.
Including boson and fermion self-energy, we obtain the full effective action at lower energy cutoff:
    \begin{eqnarray}
	S_{\text{eff}}= S_<+ \int d^4 x \bar{\psi}_{<}(-\Sigma)\psi_{<} +\frac{1}{2} \int d^4x \phi_<(-\Pi) \phi_<.
	\end{eqnarray}
After rescaling, we could obtain RG equations of all coupling constants:
	\begin{eqnarray}
	\frac{d \ln m}{dl} &=& 2z_1-1-G_2(x)\sqrt{\pi\alpha\beta} ,\\
	\frac{d \ln v}{dl} &=& 1-z_3+ G_1(x)\alpha ,\\
 	\frac{d \ln \eta}{dl} &=& 2z_1-2z_3-\frac{16}{3}\alpha+ \beta ,\\
	\frac{d \ln g^2}{dl} &=& 1-z_1-\frac{8}{3}\alpha-\frac{1}{2}\beta ,
	\end{eqnarray}
where $x\equiv \sqrt{\frac{\pi\alpha}{\beta}}=\sqrt{\frac{\Lambda\eta}{mv^2}}$ and $G_1(x)=x^2 F_1(x^2), G_2(x)=x^3 F_2(x^2)$ are two well-defined positive functions which are plotted in Fig. \ref{g_functions}.	
    \begin{figure}
	\centering
	\subfigure{\includegraphics[height=5cm]{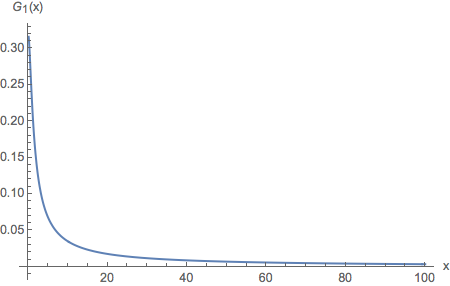}}
	\subfigure{\includegraphics[height=5cm]{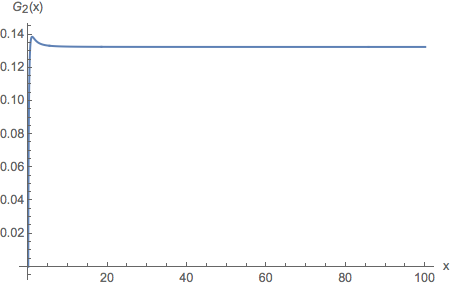}}
	\caption{The  numerical values of the $G_i$ functions are plotted. } \label{g_functions}
	\end{figure}	
Here, $z_i$ is the scaling dimensions of spatial coordinates $x_i$ which were defined in main text. We assume that the fermion mass and Fermi velocity are finite at the fixed point since they are measurable quantities defining the dispersion of fermions. Consequently, the scaling dimensions of spatial coordinates should satisfy $z_3=1+ G_1 \alpha$ and $z_1=\frac{1}{2}(1+G_2 \sqrt{\alpha \beta})$. Note that fermions are free from anomalous dimension due to instantaneous Coulomb interaction. The RG flow is best captured by the RG equations of dimensionless coupling constants $\alpha=\frac{g^2}{8\pi^2 v}$ and $\beta=\frac{g^2mv}{8\pi \Lambda\eta}$:
	\begin{eqnarray}
	\frac{d \alpha}{dl} &=& -\alpha^2[ \pi G_1(x)+\frac{16}{3}] , \\
	\frac{d \beta}{dl} &=& \beta[1-\beta+\alpha\pi  G_1(x)-\sqrt{\pi \alpha\beta} G_2(x) ]  ,
	\end{eqnarray}
where the functions $G_1$ and $G_2$ are well-defined on real axis. Moreover, we know their values at zero and infinity: $G_1(0)=\frac{1}{\pi},G_2(0)=0$ and $G_1(\infty)=0, G_2(\infty)=\frac{3}{16\sqrt{2}}$. Near $x=0$, they can be Taylor expanded as
    \begin{eqnarray}
	 G_1(x) &=&\frac{1}{\pi}-\frac{ x^2}{4}+ O(x^3) , \\
	 G_2(x) &=& \frac{1}{4\pi}(-3+4\ln 2-4\ln x)x+O(x^3).
    \end{eqnarray}

By iterating above coupled RG equations (see the RG flow diagram in the main text), it turns out that there are two fixed points with $(\alpha^\ast,\beta^\ast)=(0,0)$ and $(\alpha^\ast,\beta^\ast)=(0,1)$. The former represents the  noninteracting fixed point and is unstable. However, the latter is a stable fixed point. It is worth to point out that $\eta$ flows to zero at infrared even though the bare value of $\eta_0$ is close to one because of isotropic Coulomb interaction in the ultraviolet. This is qualitatively consistent with RPA calculations that low-energy double-Weyl fermions screen the Coulomb interactions in an anisotropic way. For simplicity, we employ the following linearized RG equations to capture the two fixed points:
    \begin{eqnarray}
	\frac{d \alpha}{ dl} &=& -\frac{19}{3} \alpha^2, \\
	\frac{d\beta}{dl} &=& (1-\beta+\alpha)\beta,
    \end{eqnarray}
from which it is clear that $(\alpha^\ast,\beta^\ast)=(0,1)$ is stable fixed point while the noninteracting one $(\alpha^\ast,\beta^\ast)=(0,0)$ is unstable. At the fixed point of noninteracting double-Weyl fermons, $\beta$ is strongly relevant while $\alpha$ is marginally irrelevant. Relevant $\beta$ around the fixed point of (0,0) drives the system away from this unstable fixed point. On the other hand, near the stable fixed point (0,1), $\alpha$ is marginally irrelevant while $\beta$ is irrelevant. Since $\alpha \propto \frac{g^2}{v}$ describes the strength of Coulomb interaction, we conclude that Coulomb interaction is marginally irrelevant in the infrared limit, which gives rise to logarithmic corrections to physical quantities as we mentioned in the main text. For the stable fixed point, because $\alpha^\ast=0$ the scaling properties of spatial coordinates are same at the tree level, namely, $z_1=\frac{1}{2},z_3=1$ at the fixed point.
	
\subsection{4. Screening of a charge impurity}
Consider the case of noninteracting double-Weyl fermions first. As mentioned in main text, induced charge density in momentum space by a Coulomb impurity is given by
	\begin{eqnarray}
	\rho(\vec q) &=& Ze V(\vec q)\Pi(\omega=0,\vec q)\nn\\
    &=&-Ze g_0^2 \frac{B_\perp q_\perp^2+B_3 |q_3|}{q_\perp^2+q_3^2},
	\end{eqnarray}
from which we can compute the density of induced charge in the real space by the Fourier transformation: $\rho(\vec r)=\int \frac{d^3 q}{(2\pi)^3} \rho(\vec q) e^{-i \vec q\cdot \vec r}$. We denote $\rho(\vec r)=\rho_\textrm{I}(\vec r)+\rho_\textrm{II}(\vec r)$, where
	\begin{eqnarray}
	\rho_\textrm{I}(r_\perp,r_3) &=& -Ze g_0^2B_\perp\frac{1}{(2\pi)^3} \int d^2 q_\perp dq_3 \frac{
    q_\perp^2}{q_\perp^2+q_3^2} e^{-iq_3 r_3-i q_\perp r_\perp} \nn\\
    &=&-Ze g_0^2B_\perp\frac{\pi}{(2\pi)^3}  \int d^2 q_\perp |q_\perp|  e^{-|r_3| |q_\perp|} e^{-i q_\perp r_\perp} \nonumber \\
	&=& -Ze g_0^2B_\perp\frac{\pi}{(2\pi)^3}  \int d q_\perp d\theta q_\perp^2 e^{-|r_3| q_\perp} e^{-i q_\perp
    |r_\perp| \cos\theta} \nn\\
    &=&-Ze g_0^2B_\perp\frac{\pi}{(2\pi)^2}  \int d q_\perp  q_\perp^2 e^{-|r_3| q_\perp} J_0(q_\perp |r_\perp|) \nn \\
	&=& -Ze g_0^2B_\perp\frac{\pi}{(2\pi)^2} \frac{-r_\perp^2+2r_3^2}{(r_\perp^2+r_3^2)^{5/2}},
	\end{eqnarray}
and
	\begin{eqnarray}
	\rho_\textrm{II}(r_\perp,r_3) &=& -Ze g_0^2B_3 \frac{1}{(2\pi)^3} \int d^2 q_\perp dq_3 \frac{
    |q_3|}{q_\perp^2+q_3^2} e^{-iq_3 r_3-i q_\perp r_\perp}\nn\\
    &=&-Ze g_0^2B_3 \frac{\sqrt{\pi}}{(2\pi)^3} \int d^2 q_\perp  G_{1,3}^{2,1} \left(\frac{q_\perp^2 r_3^2}{4} \Big| \begin{array}{cccc} 0 \\ 0,0,\frac{1}{2} \end{array} \right) e^{-i q_\perp r_\perp}   \nn\\
	&=& -Ze g_0^2B_3 \frac{\sqrt{\pi}}{(2\pi)^2} \int d q_\perp  q_\perp J_0(q_\perp |r|)G_{1,3}^{2,1}
    \left(\frac{q_\perp^2 r_3^2}{4} \Big| \begin{array}{cccc} 0 \\ 0,0,\frac{1}{2} \end{array} \right)  \nn\\
	&=& -Ze g_0^2B_3 \frac{1}{(2\pi)^2} \frac{2(\sqrt{r_\perp^2 + r_3^2}-r_3
    \sinh^{-1}(\frac{r_3}{|r_\perp|}))}{(r_\perp^2 + r_3^2)^{3/2}}.
	\end{eqnarray}
Put them together, we have induced charge density:
	\begin{eqnarray}
	\rho(\vec r) &=& \rho_\textrm{I}(\vec r)+\rho_\textrm{II}(\vec r)\nn\\
    &=& -Ze g_0^2 \frac{1}{(2\pi)^2}\left(\pi B_\perp \frac{-r_\perp^2+2r_3^2}{(r_\perp^2+r_3^2)^{5/2}}+ 2B_3\frac{\sqrt{r_\perp^2 + r_3^2}-r_3 \sinh^{-1}(\frac{r_3}{|r_\perp|})}{(r_\perp^2 + r_3^2)^{3/2}}\right).
	\end{eqnarray}
Partially induced charge densities introduced in the main text are given by
	\begin{eqnarray}
	Q_\perp(r_\perp) &=& \int d r_3 \rho(r_\perp,r_3) \nn\\
    &=& -Ze g_0^2 \int d r_3 \int \frac{d q_3}{2\pi} \frac{d
    q^2_\perp}{(2\pi)^2} \frac{B_\perp q_\perp^2+B_3 |q_3|}{q_\perp^2+q_3^2} e^{-i q_\perp \cdot r_\perp} e^{-i q_3 r_3}   \nn\\
	&=& -Ze g_0^2B_\perp   \int \frac{d q^2_\perp}{(2\pi)^2} \frac{q_\perp^2}{q_\perp^2} e^{-i q_\perp \cdot
    r_\perp} \nn\\ &=& -Zeg_0^2 B_\perp \delta^{(2)} (r_\perp),  \\
	Q_3(r_3) &=& \int d^2 r_\perp\rho(r_\perp,r_3) \nn\\
    &=&-Ze g_0^2 \int d^2 r_\perp \int \frac{d q_3}{2\pi} \frac{d q^2_\perp}{(2\pi)^2} \frac{B_\perp q_\perp^2+B_3 |q_3|}{q_\perp^2+q_3^2} e^{-i q_\perp \cdot r_\perp} e^{-i q_3 r_3}   \nonumber \\
	&=& -Ze g_0^2B_3  \int \frac{d q_3}{2\pi} \frac{1}{|q_3|} e^{-i q_3 r_3} \nn\\
    &=&Ze g_0^2B_3 \frac{\gamma+\log(|r_3|)}{\pi},
	\end{eqnarray}
where $\gamma$ is Euler-Gamma constant.

We now consider the case of interacting double-Weyl fermions. In a similar way, the induced charge density in momentum space is given by
	\begin{eqnarray}
	\rho(\vec q)&=& Zeg^2 V(\vec q) \Pi(\omega=0,\vec q) \nn\\
    &=&-Zeg^2 \frac{B_\perp q_\perp^2+B_3 |q_3|}{q_\perp^2+|q_3|},
	\end{eqnarray}
from which we obtain partially-integrated induced charge density in real space:
	\begin{eqnarray}
	Q_\perp(r_\perp) &=& -\int dr_3 \int \frac{d^3 q}{(2\pi)^3} Zeg^2\frac{B_\perp q_\perp^2+B_3 |q_3|}{q_\perp^2+|q_3|}
    e^{-i q \cdot r} \nn\\
    &=& -Zeg^2B_\perp \delta^{(2)}(r_\perp), \\
	Q_3(r_3) &=& -\int d^2 r_\perp  \int \frac{d^3 q}{(2\pi)^3} Zeg^2\frac{B_\perp q_\perp^2+B_3 |q_3|}{q_\perp^2+|q_3|}
    e^{-i q \cdot r} \nn\\
    &=& -Zeg^2B_3 \delta(r_3).
	\end{eqnarray}
As the total induced charge should not depend on the order of integration, i.e., $\int d^2 r_\perp Q_\perp(r_\perp)= \int dr_3 Q_3(r_3)$, it is required that $B_\perp = B_3$. So, induced charge density in real space is $\rho(\vec r)= -Zeg^2 B_3 \int  \frac{d^3 q}{(2\pi)^3} \frac{q_\perp^2+|q_3|}{q_\perp^2+|q_3|} e^{-i \vec q\cdot \vec r} =  -Zeg^2 B_3\delta^{(3)}(\vec r)$.

\subsection{5. Specific heat}
Since Coulomb interaction is marginally irrelevant at the stable fixed point, we expect logarithmic corrections to physical quantities. Here, we specifically calculate how specific heat of double-Weyl fermions are affected by Coulomb interactions in low temperature. Corrections to other measurable quantities may be obtained similarly by taking account of scaling hypothesis near the fixed point as well as the flow of coupling constants. According to scaling hypothesis, free energy density scales as $f=b^{-2z_1-z_3-1} f_0$, where $b= \frac{\Lambda}{T}$ and $z_1,z_3$ are scaling exponent of coordinates along $x$ (or $y$) and $z$ directions respectively. As a result, specific heat $C=-T\frac{\partial^2 f}{\partial T^2}$ is given by
	\bea
	C(T)= b^{-2z_1-z_3} C_0(\Lambda),
	\eea
from which we obtain the RG equation of specific heat as follows:
	\bea
	\frac{d C(l)}{d l}= -(2z_1+z_3)C(l) \label{diff_specific_heat} ,
	\eea
where $l \equiv \ln b$. Near the fixed point, $z_1=\frac{1}{2}(1+G_2 \sqrt{\alpha \beta})$ and $z_3=1+G_1 \alpha$ (see the preceding section of this Appendix for details). Because $\alpha$ is magically irrelevant and $\beta$ is irrelevant near the stable fixed point, the RG flow along $\beta$ direction is much faster than that along $\alpha$ direction. This is obviously seen in Fig. 2(b) of the main text. As logarithmic correction is caused by marginal flow of $\alpha$, we can set $\beta=\beta^\ast=1$ in the following calculations as a good approximation. For $\alpha$ close to $\alpha^\ast=0$, we obtain the asymptotic behavior of $G_1$ and $G_2$:
	\bea
	G_1&\approx&\frac{1}{\pi}-\frac{\pi}{4}\alpha, \label{G1} \\
	G_2&\approx&-c_1\sqrt{\alpha}-\sqrt{\frac{\alpha}{4\pi}}\ln \alpha ,\label{G2}
	\eea
where $c_1>\frac{1}{\pi}$ is a positive constant. Note that the logarithmic term in $G_2$ is due to fermion self-energy. By plugging Eq. (\ref{G1}), Eq. (\ref{G2}), and Eq. (9) of the main text into Eq. (\ref{diff_specific_heat}), we obtain:
	\bea
	\frac{d \ln C}{d l}= -2-c_2\alpha(l)+ \frac{1}{\sqrt{4\pi}} \alpha(l)\ln\alpha(l),
	\eea
where $\alpha(l)= \frac{\alpha_0}{1+\frac{19\alpha_0}{3}l}$ and $c_2= c_1-\frac{1}{\pi} \approx 0.037$. The first term will bring the usual logarithmic correction to specific heat as we expect. However, the second term becomes more dominant in the infrared or at low enough temperature. The deviation of specific heat from the usual Fermi liquid behavior is apparent:
	\bea
	C(T) &\propto& T^2 \exp\left[ -\frac{3c_2}{19}\ln(1+\frac{19\alpha_0}{3}\ln \frac{T_0}{T} )-
    \frac{3}{76\sqrt{\pi}}\left( \ln(1+\frac{19\alpha_0}{3}\ln \frac{T_0}{T} )\right)^2 \right] \nn\\
	&=& \frac{T^2}{(1+\frac{19\alpha_0}{3}\ln \frac{T_0}{T})^{\frac{3c_2}{19}}} \times \exp\left[
    -\frac{3}{76\sqrt{\pi}}\left( \ln(1+\frac{19\alpha_0}{3}\ln \frac{T_0}{T} )\right)^2 \right], \label{specific_heat_1}
	\eea
The exponential factor in the last line of Eq. (\ref{specific_heat_1}) is a new phenomenon of marginal Fermi liquids we find in double-Weyl fermion (the usual logarithmic corrections appear without the second exponential factor in Eq. (\ref{specific_heat_1})). Fig. (S2) shows the schematic plot of specific heat. The solid line is plotted from Eq. (\ref{specific_heat_1}) and is more exotic comparing to the dashed line which is the usual logarithmic corrections appearing near common marginally-irrelevant fixed point. The origin of this more exotic corrections lies in the unusual logarithmic contribution in $G_2$, which comes from the interplay between the anisotropic spectrum of double-Weyl fermions and isotropic bare Coulomb interactions.

\begin{figure}
\includegraphics[width=8cm]{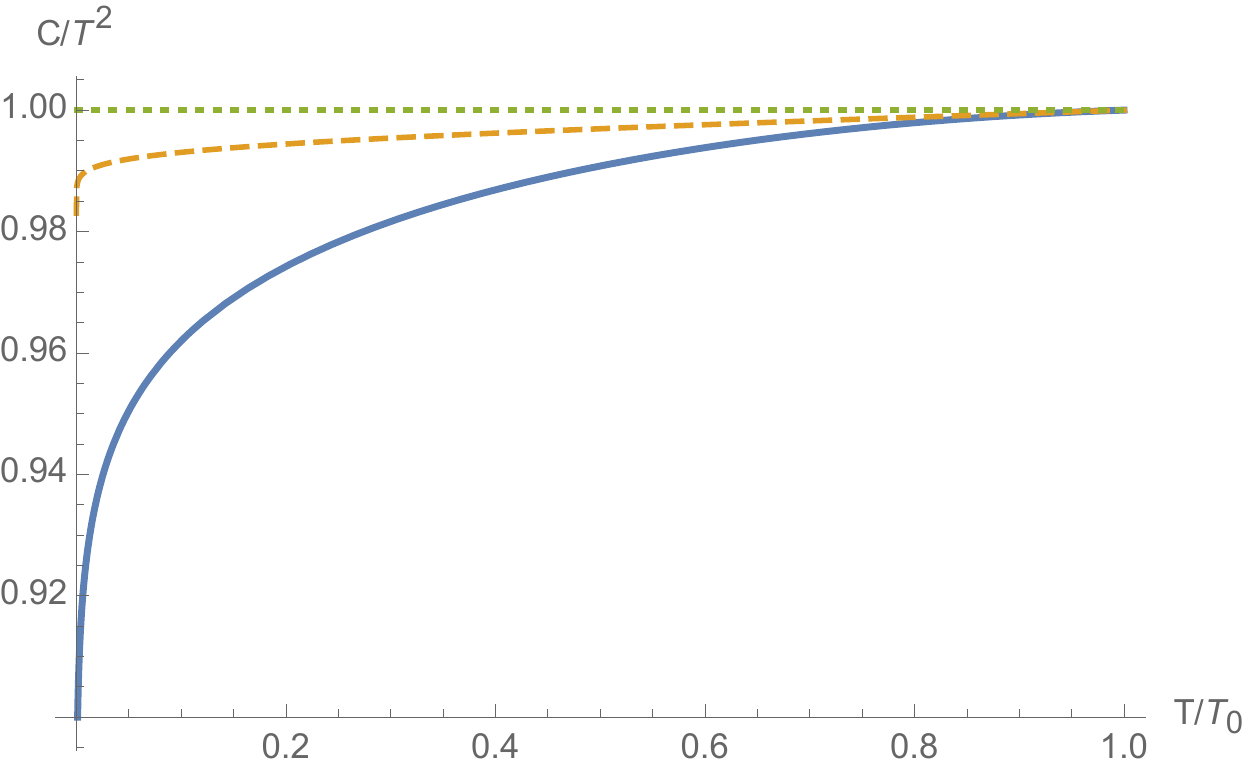}
\caption{Schematic plot of specific heat in double-Weyl semimetals with and without Coulomb interactions. The dotted line is the specific heat from electronic contribution in noninteracting double-Weyl fermions.  The solid line is plotted from Eq. (\ref{specific_heat_1}) [or Eq. (10) in the main text]. The dashed line is plotted from the same equation but without the exponential factor.}
\end{figure}

\end{widetext}

\end{document}